\documentclass[preprint2]{aastex6}

\usepackage{natbib}
\usepackage{gensymb}

\slugcomment{Accepted for publication in the Astronomical Journal}

\shorttitle{The new eclipsing CV J1923}
\shortauthors{Kennedy et al.}

\begin{document}

\title{The new eclipsing CV MASTER OTJ192328.22+612413.5 - a possible SW Sextantis Star}

\author{M. R. Kennedy\altaffilmark{1}$^{,}$\altaffilmark{2}, P. Callanan\altaffilmark{1}, P. M. Garnavich\altaffilmark{2}, P. Szkody\altaffilmark{3}, S. Bouanane\altaffilmark{1}, B. M. Rose\altaffilmark{2}, P. Bendjoya\altaffilmark{4}, L. Abe\altaffilmark{4}, J. P. Rivet\altaffilmark{4} and O. Suarez\altaffilmark{4}}

\altaffiltext{1}{Department of Physics, University College Cork, Cork, Ireland}
\altaffiltext{2}{Department of Physics, University of Notre Dame, Notre Dame, IN 46556}
\altaffiltext{3}{Department of Astronomy, University of Washington, Seattle, WA}
\altaffiltext{4}{Laboratoire Lagrange UMR 7293, Universit\'{e} de Nice Sophia-Antipolis, Observatoire de la C\^{o}te d'Azur, France}
\email{markkennedy@umail.ucc.ie}

\begin{abstract}
We present optical photometry and spectroscopy of the new eclipsing Cataclysmic Variable MASTER OTJ192328.22+612413.5, discovered by the MASTER team. We find the orbital period to be $P=0.16764612(5)$ day /$4.023507(1)$ hour. The depth of the eclipse ($2.9\pm0.1$ mag) suggests that the system is nearly edge on, and modelling of the system confirms the inclination to be between $81.3-83.6\degree$. The brightness outside of eclipse varies between observations, with a change of $1.6\pm0.1$ mag. Spectroscopy reveals double-peaked Balmer emission lines. By using spectral features matching a late M-type companion, we bound the distance to be $750\pm250$ pc, depending on the companion spectral type. The source displays 2 mag brightness changes on timescales of days. The amplitude of these changes, along with the spectrum at the faint state, suggest the system is possibly a dwarf nova. The lack of any high excitation HeII lines suggests this system is not magnetically dominated. The light curve in both quiescence and outburst resembles that of Lanning 386, implying MASTER OTJ192328.22+612413.5 is a possible cross between a dwarf nova and a SW Sextantis star. 
\end{abstract}


\section{Introduction}
Binary systems in which a main or quasi-main sequence star transfers material onto the surface of a white dwarf (WD; the primary star) are known as cataclysmic variables (CVs). There are 2 main categories for CVs: strongly magnetic and weakly magnetic systems. In systems containing a WD with a weak magnetic field, the material lost by the secondary leaves the L1 point to stream toward the WD and forms an accretion disk around the WD. Once the disk is formed the stream strikes it and forms a bright (or hot) spot. The disk can extend down to the surface of the WD, and the material accretes onto the WD. In magnetic CVs, the gas stream leaving the L1 point follows the magnetic field lines to the WD and forms an accretion column for WDs with strong magnetic fields (also known as polar systems), or else for WDs with moderate magnetic fields (known as intermediate polars, or IPs), the material forms a disk until the magnetic pressure overcomes the ram pressure, at which case the disk truncates and the material flows to the WD in an accretion curtain. CVs have several photometric classifications depending on the light curves of their outbursts - classical novae  have one observed large amplitude ($\Delta m > 6$) outburst; dwarf novae  typically have repeated, smaller amplitude ($2<\Delta m<6$) outbursts; and nova-likes have no observed outbursts, but sometimes transition between high and low states of accretion \citep{Warner1995}.

SW Sextantis stars are a subclass of high accretion rate CVs. Their defining features are: (1) they are nova-like CVs, with an orbital period $3<P<4$ hours; (2) they show He II $\lambda$4686 emission with a strength of half $H\beta$ or larger; (3) the radial velocities of the emission lines vary periodically, with the Balmer lines lagging behind the expected phase for a WD based on the eclipse; (4) absorption lines are only visible for a part of the orbit opposite the eclipse phase \citep{Thorstensen1991}. The emission lines of SW Sextantis stars are also single-peaked lines, making candidates easy to identify. The single-peaked emission lines are thought to originate from material encountering a magnetic accretion curtain close to the surface of the white dwarf \citep{Hoard2003}. SW Sextantis stars are now thought to be the dominant population of CVs with periods between 3 and 4 hours \citep{Rodriguez-Gil2007}. Several SW Sex stars show circular polarization, indicative of a magnetic nature \citep{Rodriguez-Gil2001}.

Lanning 386 is an eclipsing CV with a period of 3.94 hours which defies easy classification \citep{Brady2008}. The system displays recurring outbursts with an amplitude of $\Delta m \approx 2$, but is at quiescence most of the time. This would lead Lanning 386 to be classified as a dwarf nova. The spectrum of Lanning 386 in quiescence is consistent with this classification, but the spectrum in outburst tells another story. The spectrum in quiescence shows single-peaked Balmer emission lines, and in outburst displays strong, single-peaked He I and He II lines, and the strong excitation C IV line, consistent with a SW Sextantis star \citep{Groot2000}. However, in quiescence and outburst, Lanning 386 doesn't show the emission line radial velocity phase lag expected of the SW Sextantis stars \citep{Brady2008}. Also of interest is the recurrence time of the outbursts, which is only a few days. Other scenarios have been considered, such as Lanning 386 being a VY Scl system (which are nova-like, non-magnetic or weakly magnetic CVs which exhibit low states on a recurrence time of years \citep{Leach1999}) or even an IP, but again, Lanning 386 only shares some of the characteristic features of these classes, never all. Its true classification still remains a mystery, but is currently best described as a dwarf nova in quiescence and a SW Sextantis star in outburst.

MASTER OTJ192328.22+612413.5 (hereafter referred to as J1923), the subject of this paper, was initially discovered by the MASTER-Tunka auto-detection system in April 2014 (Balanutsa et al., ATel 6097). The magnitude of J1923 was observed to be 19.2 in 2010, but had brightened to an unfiltered magnitude of 17.5 on April 20, 2014 and 17.7 on April 22nd, 2014. It was classified as a CV due to its 2 magnitude difference with respect to reference images. There is also a UV counterpart to J1923, GALEX J192328.3+612413, which has a FUV magnitude of $19.42\pm0.15$ and NUV of $19.68\pm0.10$, obtained on June 1, 2007. J1923 was also detected in the infared as WISE J192328.25+612413.1. The ALLWISE Source Catalog lists magnitudes of $W1=16.58\pm0.05$ and $W2=16.6\pm0.1$ for J192328.25+612413.1. It was also suggested that J1923 might be a magnetically dominated CV, or polar, due to its proximity to the ROSAT x-ray source 1 RXS J192333.2+612507, which has a large error circle of 14".

In this paper, we present photometry and spectroscopy of the newly discovered CV J1923, and use the data to obtain a period for the system and classify its CV type. We also estimate the inclination of the system based on analysis of the depth of the eclipse. Finally, we present a distance estimate using constraints on the companion spectral type.

\section{Observations}

\subsection{Photometry}
Photometry of J1923 was taken in the SDSS g band using the 1.8m Vatican Advanced Technology Telescope (VATT) located at Mount Graham International Observatory over 4 consecutive nights, starting April 29 2014 (UT). The images were taken using the VATT4K CCD with a typical exposure time of 30s. A total of 880 images were taken over these 4 nights. There are deep eclipses visible on the 3rd and 4th nights of the run. On the 1st night, there is an eclipse that is not well determined due to the onset of twilight.

A single night of V band photometry was acquired at the VATT on the night of June 28 2014 (UT). A total of 518 images were taken with a typical exposure time of 15s. There is a clear eclipse visible in these data.

Data were also acquired from the dual 1m telescopes C2PU facility at the Calern Observatory (Observatoire de la C\^{o}te d'Azur, France) in the second half of August 2014. A total of two hundred and forty five 40s exposures with no filter, one hundred 60s exposures with no filter, and eighty three simultaneous 60s exposures using V and R filters were obtained. There are eclipses visible in all of these data sets.

\begin{figure}[h!]
\epsscale{1}
\plotone{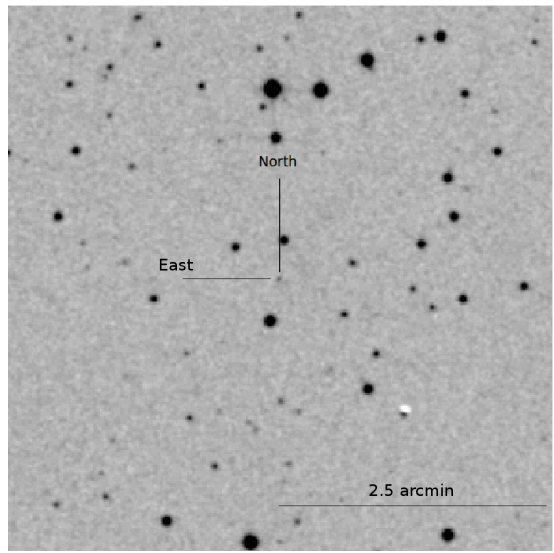}
\caption{Finding chart for J1923 +6124, taken from the STScI Digitized Sky Survey. The photometry was calibrated using the bright star just to the south of J1923, which has a g band magnitude of 15.64, V band magnitude of 15.19 and R band magnitude of 15.13.}
\label{findingchart}
\end{figure}

\subsection{Spectroscopy}
\begin{figure*}
\epsscale{1}
\plotone{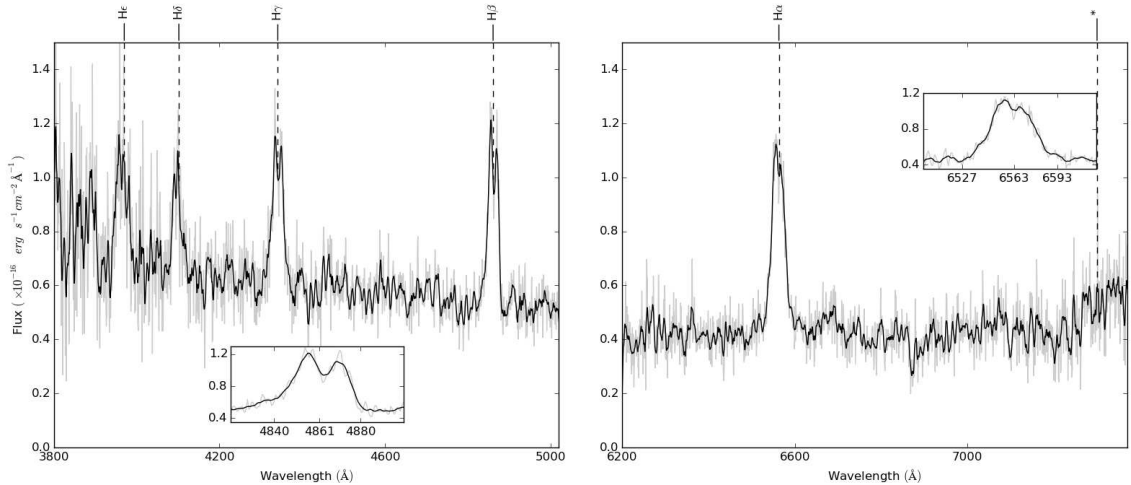}
\caption{Average spectrum of J1923 in quiescence. The left panel shows the blue spectrum while the right panel shows the red spectrum. The spectra have been boxcar smoothed with a width of 6 \AA. The grey spectrum is the unsmoothed data. The most prominent emission features are marked. The inset on the left shows the double-peaked H$\beta$ line, while the right inset shows the double peaked H$\alpha$ line. There is a clear rise towards the red in the spectrum on the right which is marked with a *, which matches a late M type star. Analysis of the lines can be seen in Table \ref{emissionproperties}.}
\label{average_spectrum}
\end{figure*}

\begin{deluxetable*}{ccccccc}
\tabletypesize{\small}
\tablecolumns{7}
\tablewidth{0pc}
	\tablecaption{Summary of Observations\label{eclipseno}}
	\tablehead{
		\colhead{Frame in}	&\colhead{Date Observed}	&\colhead{Filter}	&\colhead{State}	&\colhead{Eclipse}	&\colhead{Mid-eclipse Time}	&\colhead{Uncertainty}\\
		\colhead{Figure \ref{lightcurve}} &\colhead{2014}	&\colhead{}	&\colhead{}	&\colhead{Number}	&\colhead{(HJD-2,450,000.0)}	&\colhead{(day)}
		}
		
		\startdata
		a)	& April 28th		& SDSS-g		& Low	& -11	& 6776.992	& 0.001  \\
		b)	& April 29th		& SDSS-g		& Low	& N/A	& 	&   \\
		c)	& April 30th		& SDSS-g		& High	& 0	& 6778.83667	& 0.00009 \\
		d)	& May 1st		& SDSS-g		& High	& 6	& 6779.84257	& 0.00004 \\
		e)	& June 27th		& V			& High	& 346	& 6836.84301	& 0.00004 \\
		f)	& August 19th	& Clear		& High	& 660	& 6889.48331	& 0.00001 \\
		g)	& August 25th	& Clear		& Low	& 696	& 6895.51890	& 0.00004 \\
		h)	& August 27th	& V/R		& Low	& 708	& 6897.53010	& 0.00007 \\
		\enddata
\end{deluxetable*}

Spectroscopy of J1923 was taken using the Dual Imaging Spectrograph (DIS) on the 3.5m ARC telescope at Apache Point Observatory on the night of June 21 2014 (UT). The blue channel used the B1200 grating, which has a pixel scale of 0.62 \AA$/$pix, while the red channel used the R1200 grating, which has a pixel scale of 0.58 \AA$/$pix.  A total of five 900s long exposures were obtained on the blue and red channels. The first 3 spectra were obtained at an airmass greater than 1.4, while the final 2 spectra were taken later in the night, at an airmass of 1.2.

Figure \ref{average_spectrum} shows the average spectrum for J1923, with variations due to the orbital motion of the system unaccounted for. The average spectrum shows distinct double-peaked emission lines, corresponding to $H\alpha$, $H\beta$ and $H\gamma$ lines, along with a rise towards the red end of the spectrum. The flux of J1923 in the V band section of the spectrum in Figure \ref{average_spectrum} is $\approx0.45\times10^{-16}\;erg \: cm^{-2}\:s^{-1}\:\AA^{-1}$. This corresponds to a V band magnitude of 19.8, which is the magnitude of J1923 in the low state.

\section{Results}
\begin{figure*}
\epsscale{1.2}
\plotone{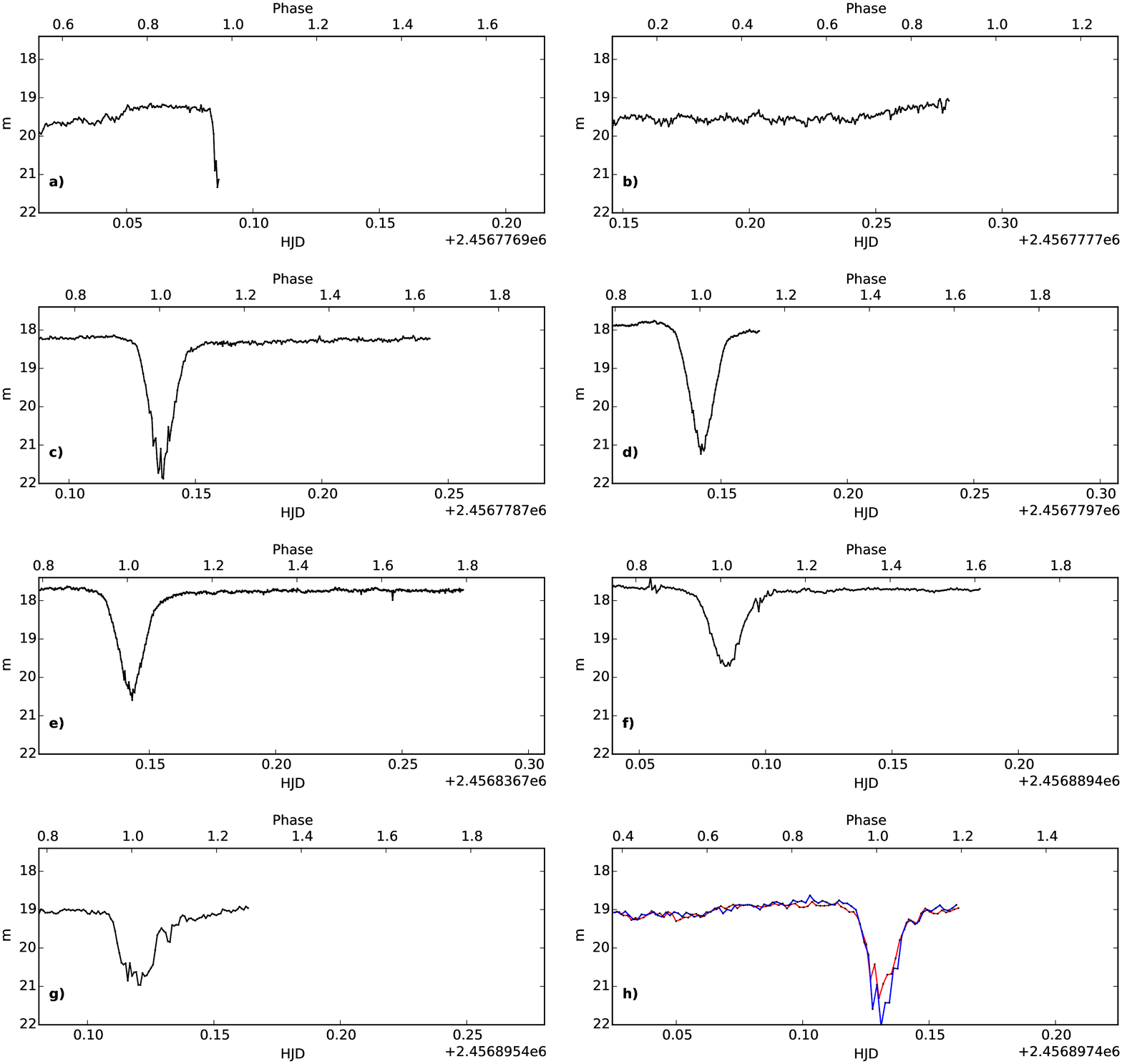}
\caption{The light curve of J1923. Frames a), b), c) and d) were taken in sequential nights in the SDSS-g band. Frames a) and b) show the light curve in quiescence, with the hump from the hot spot visible in frame a) and b) and the QPOs visible in frame b). The system went into outburst between frames b) and c), with the brightness of the system again increasing between frames c) and d). Frame e) was taken 2 months later, at the end of June 2014, in the V band with a typical exposure time of 15s. Frame f) was taken 2 months later again, at the end of August 2014, without a filter, and with a typical exposure time of 40s. Frame g) was also taken in August, with no filter and a typical exposure time of 60s as the system was in quiescence. Frame h) shows the simultaneous V and R filter data taken in August, with a typical exposure time of 60s. The V band data are displayed in blue, and the R band data are in red.}
\label{lightcurve}
\end{figure*}

The light curves for each night after standard photometric calibration can be seen in Figure \ref{lightcurve}. All data reduction and photometry was done using standard \texttt{IRAF}\footnote{\texttt{IRAF} is distributed by the National Optical Astronomy Observatory, which is operated by the Association of Universities for Research in Astronomy (AURA) under cooperative agreement with the National Science Foundation} tasks. Photometry was carried out on J1923 using the \texttt{PHOT} command from the \texttt{DIGIPHOT} package, and the images were calibrated using nearby stars from the SDSS catalogue.

\subsection{Period and Ephemeris}
\begin{figure}[h]
\epsscale{1}
\plotone{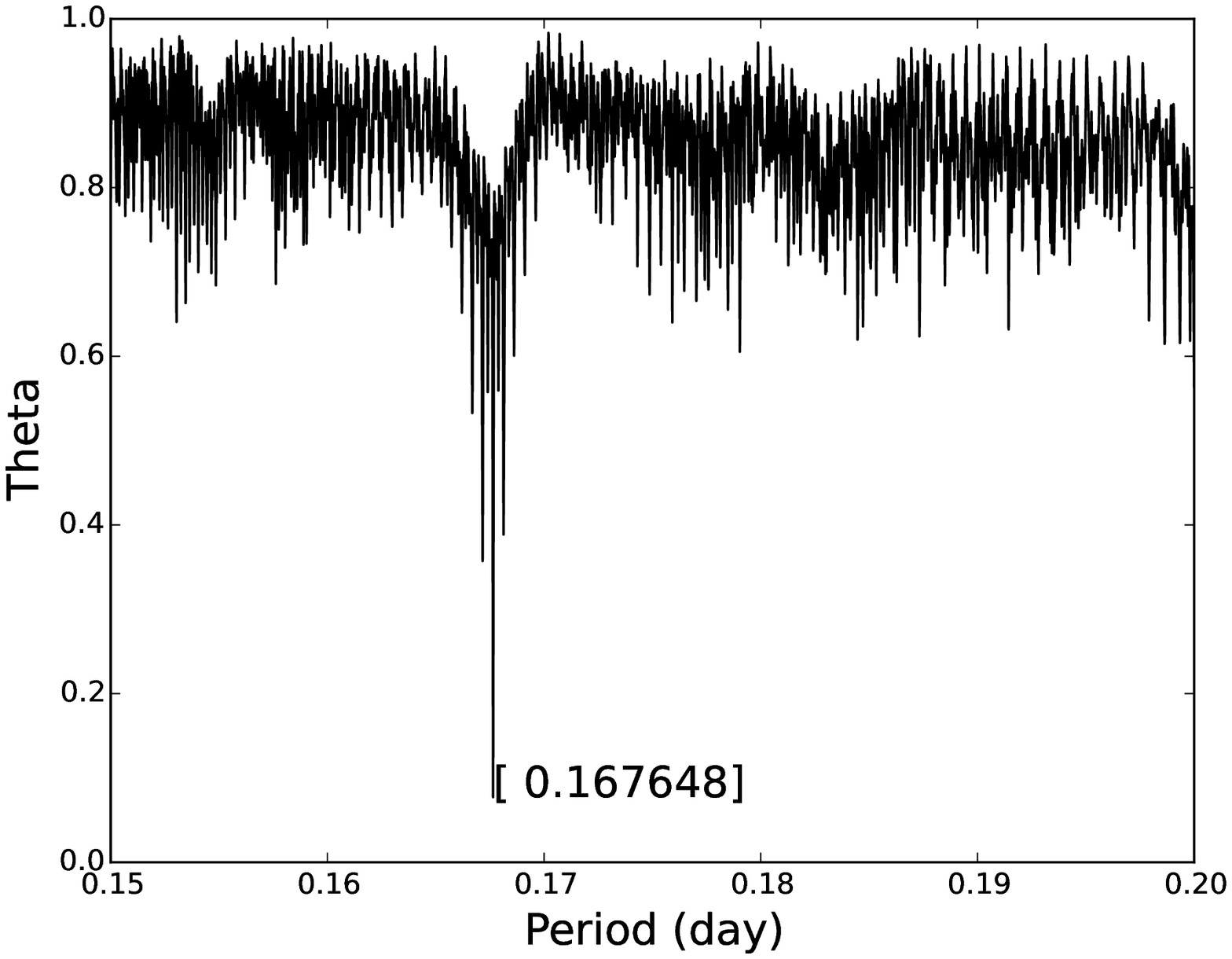}
\caption{PDM of the light curve of J1923. The deepest minimum occurs at $P=0.16765 \pm 0.00004$ day.}
\label{PDM}
\end{figure}
The period of the system was found by applying a Phase Dispersion Minimisation (PDM) \citep{Stellingwerf1978} to the data, which were first normalized to have a common non-eclipse magnitude. The period range was set to between 0.15 and 0.2 day. The resulting theta plot can be seen in Figure \ref{PDM}. The best fit period is $0.16765 \pm 0.00004$ day, with the error being the width of this peak. The periods on either side of the deepest trough were tested, and were found to give poor fits.

\begin{figure}
\epsscale{1}
\plotone{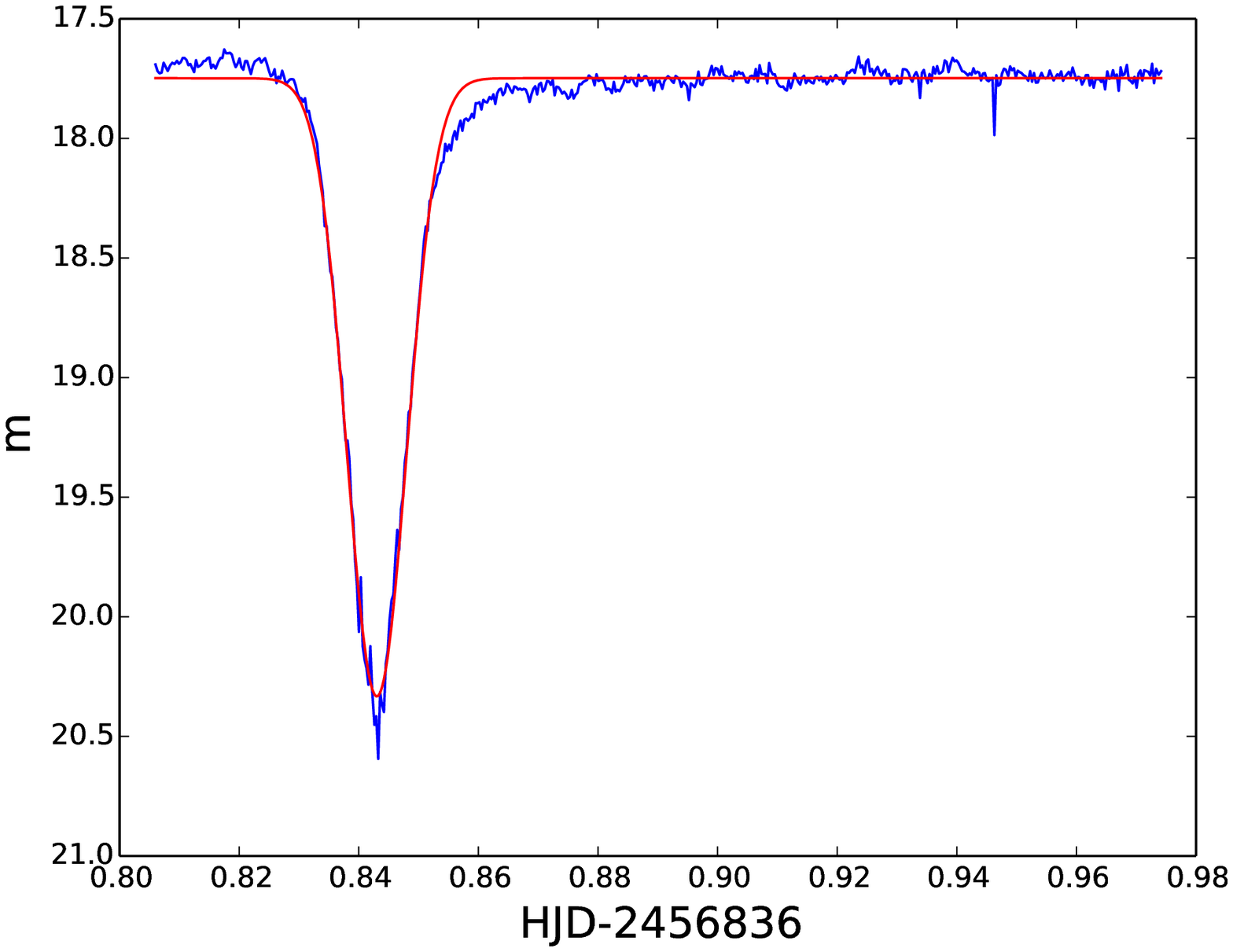}
\caption{The eclipses were modelled using a gaussian function to accurately determine the time of mid-eclipse. Above is a single nights data from the VATT from June 2014, with the red line representing the best Gaussian fit.}
\label{Gauss_Fit}
\end{figure}

The time of the midpoint of each observed eclipse was found by fitting a Gaussian to the eclipse profiles (see Figure \ref{Gauss_Fit} for an example), and using the center of the Gaussian as the time of mid-eclipse, with the error in the time being the error in fitting the center of the Gaussian. These times and the associated eclipse numbers can be seen in Table \ref{eclipseno}, where the eclipse number was found by taking the 3rd night's eclipse as 0, and using the newly discovered period to find the remaining eclipse numbers relative to this night. The first nights eclipse has a much higher error than other nights as the eclipse was only partially observed. For this night, a model eclipse was constructed using a Gaussian, and the midpoint adjusted manually to see where the Gaussian matched the beginning of the eclipse.

\begin{figure*}
\epsscale{1}
\plotone{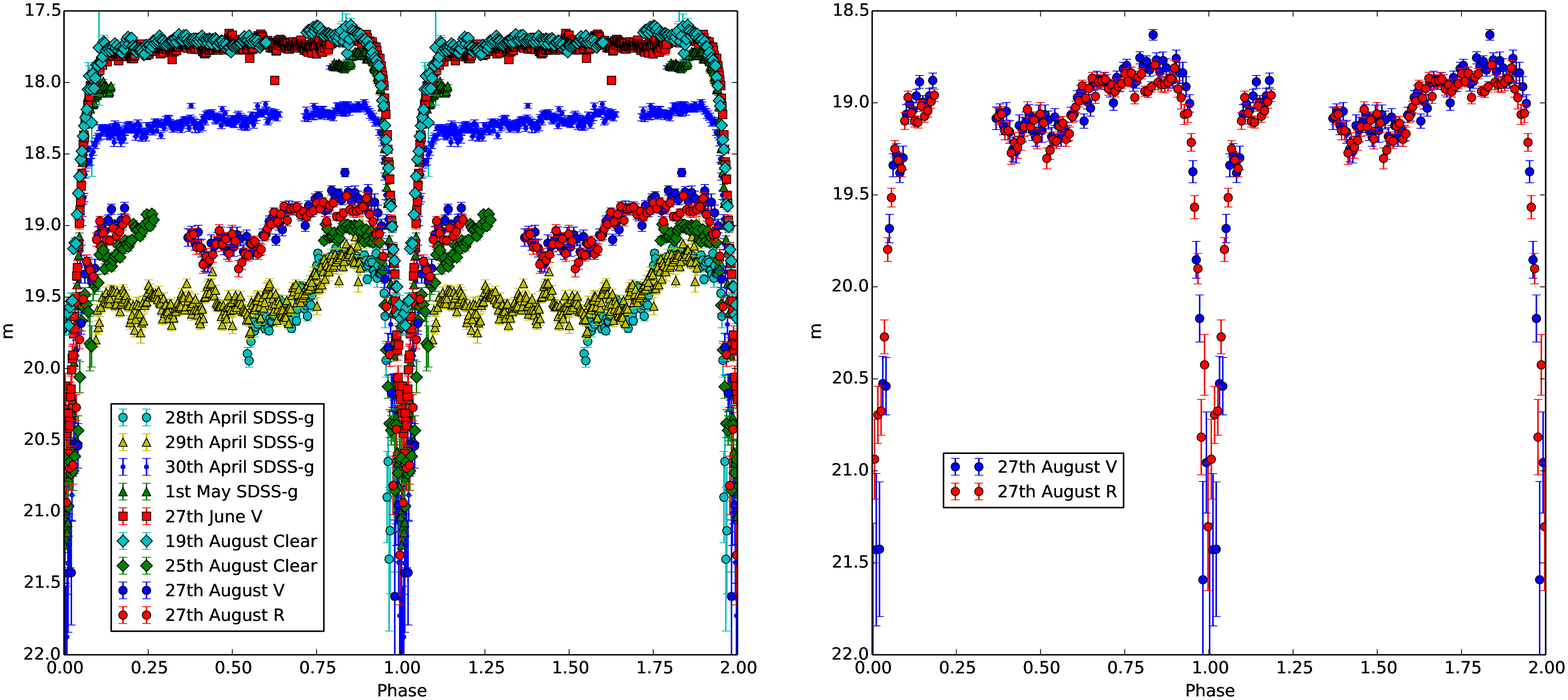}
\caption{The left panel shows the phased light curve for J1923 for all 8 nights data using the linear ephemeris. The right panel shows just the simultaneous R and V band data taken in August, 2014}
\label{phased_lc}
\end{figure*}

A linear fit was applied to the data, and the resulting ephemeris was 
\begin{equation}
T_{mid} (HJD) =2456778.83690(3)+0.16764612(5)E
\label{lineareph}
\end{equation}
The phased light curve using the linear ephemeris can be seen in Figure \ref{phased_lc}.

\subsection{Light curve Morphology}
\begin{figure}
\epsscale{1}
\plotone{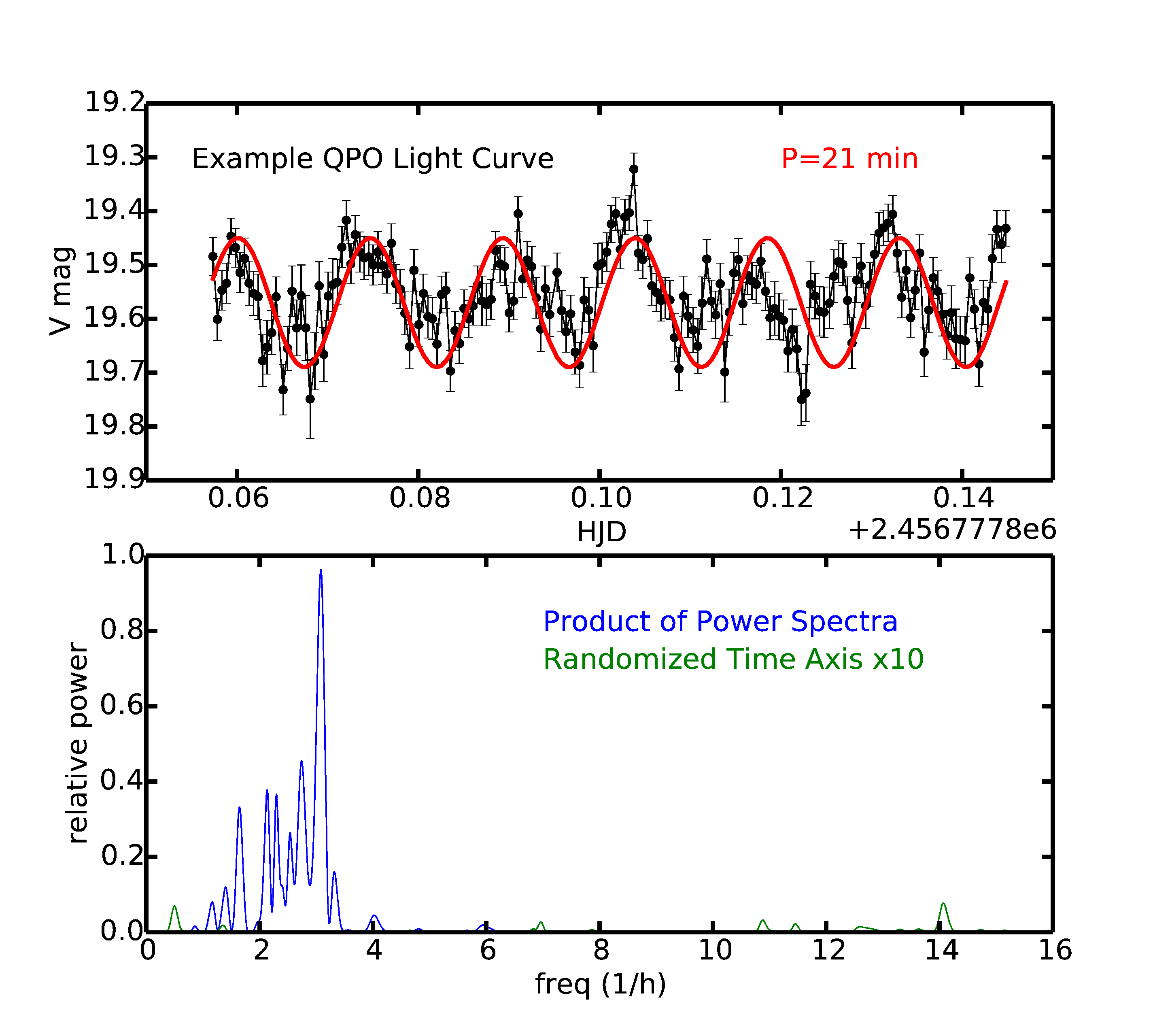}
\caption{The bottom panel shows the power spectrum of the data with the eclipses masked, in order to look for the period of the QPOs. The blue shows the power spectrum of the normal data, while the green shows the power spectrum of the data after it had been randomised, to make sure the peaks in the original power spectrum were not due to noise. The top panel shows a sine wave with a frequency of 3.32 h$^{-1}$ plotted alongside the data from 2014 April 29, when the system was in its low state and had the most prominent QPOs.}
\label{QPO}
\end{figure}

The first 2 nights of observations show a light curve with a pre-eclipse hump starting at phase 0.7, along with quasi-periodic oscillations (QPOs) just after the eclipse. The data were masked to remove the eclipses, and then each night of data was subjected to a Lomb-Scargle Periodogram (\citealt{Lomb76}; \citealt{scargle82}). The power spectrum for each night were then multiplied together to reduce the power of the noise peaks while maintaining power in the peaks common to each data set. The resulting power spectrum can be seen in the lower half of Figure \ref{QPO}. There are several strong peaks around a frequency of $\sim$3.3 hour$^{-1}$ (P$_{QPO}$ $\sim20$ minute). The top panel of Figure \ref{QPO} shows a sine wave with a period of 21 minute plotted on top of the data from 2014 April 29, when the system was in its low state and the QPOs were most prominent. The broad power spectrum confirms these oscillations are not coherent and cannot be solely due to the spin of the central WD. Kilosecond QPOs have been seen in SW Sextantis stars before, and are thought to arise from magnetic white dwarfs that are drowned by a high accretion rate \citep{Patterson2002}.

The light curve also shows 2 distinct states. The quiescent state has a non-eclipse magnitude of 19.5 in the g band from the first 2 nights (frames a) and b) of Figure \ref{lightcurve}), and the pre-eclipse hump and QPOs are distinct. The system had an out-of-eclipse magnitude that gradually decreased from night to night as the system went into outburst. On the first night of outburst (frame c)), the average magnitude was 18.2, while on the second night of outburst, the system had brightened again to $m < 18$. The depth of the eclipse in outburst, which was well covered by observations, was $\Delta m=3.2\pm0.1$ mag as seen in the g band and $\Delta m =2.9\pm0.1$ in the V band.

The eclipse width was found by fitting a Gaussian to the V-band data, after conversion from magnitude to flux. The resulting fit gave an eclipse width (FWHM) of $0.081\pm0.001$ in phase.

\subsection{Emission Spectrum}
By using the calculated ephemeris, we found the phase of each of the spectra obtained: 0.81, 0.87, 0.21, 0.29 and 0.35. The rise towards the red end of the spectrum is thought to be molecular bands visible from the companion star which match that of a late M type star, similar to the companion of Lanning 386 \citep{Brady2008}. This is further discussed in Section \ref{Model_Spec_Sec}. The equivalent width (EW) of the $H\alpha$, $H\beta$ and $H\gamma$ lines can be seen in Table \ref{emissionproperties}, along with the Balmer decrement.

\begin{deluxetable}{lccccc}
\tabletypesize{\scriptsize}
\tablecolumns{6}
\tablewidth{0pc}
	\tablecaption{Properties of the emission lines seen in the average spectrum of J1923. \textit{(a)The FWZIs are given as lower limits since the ability to determine the extension of the line wings is usually set by the signal-to-noise ratio in the data} \textit{(b) Measured using a Gaussian fit}\label{emissionproperties}}
	\tablehead{
		\colhead{Feature}	&\colhead{E.W.}	&\colhead{Flux}	&\colhead{FWZI$^{a}$}	&\colhead{FWHM$^{b}$}	&\colhead{Balmer}\\
		\colhead{} &\colhead{}	&\colhead{}	&\colhead{}	&\colhead{}	&\colhead{Decrement} \\
		\colhead{} &\colhead{\AA}	&\colhead{$10^{-17} erg \: cm^{-2}\:s^{-1}$}	&\colhead{km s$^{-1}$}	&\colhead{km s$^{-1}$}	&\colhead{Decrement}
		}
		
		\startdata
		$H\alpha$ & 47 & 217 & $\gtrsim2800$ & 1500 & 1.20 \\
		$H\beta$	 & 35 & 181 & $\gtrsim3000$ & 1700 & 1\\
		$H\gamma$ & 24 & 146 & $\gtrsim3800$ & 2000 & 0.81\\
		\enddata
\end{deluxetable}

The full width at zero intensity (FWZI) of the $H\alpha$, $H\beta$ and $H\gamma$ was measured within \texttt{IRAF}, as were the peak-to-peak separations of the $H\alpha$, $H\beta$ and $H\gamma$ lines, found by fitting a double Lorentzian to the line region, and can be seen in Table . The fits gave separations of 16\AA, 16\AA$\:$ and 15\AA$\:$ respectively. The asymmetry between the blue and red peaks could be due to the presence of the hotspot.

\section{Discussion}

\subsection{Inclination and Mass Ratio of J1923}
\label{inclination_sec}
The depth of the eclipse is $3.2\pm0.1$ mag seen in the g band on night 3 and night 4 of the data and $2.9\pm0.1$ mag in the V band on night 5. The majority of CVs with an eclipse depth close to or larger than 2.5 mag have an inclination angle greater than 85 degrees \citep{Ritter2003}. Hence, the inclination of J1923 is expected to be $> 85^{o}$, suggesting that J1923 is a nearly edge on disk system.

The $H\alpha$ peak-to-peak separation of 16 \AA$\:$ corresponds to a velocity separation of $800\:\:km\:s^{-1}$. This would normally limit the inclination of the system to be $< 30^{o}$ \citep{Horne1986}. However, such a deep eclipse is hard to explain with such a low inclination. Instead, it is possible that this low peak-to-peak separation is due to a truncation of the accretion disk far from the surface of the primary, which might be due to the presence of a strong magnetic field on the WD, or a low velocity component (spot or wind perpendicular to the disk or magnetic curtain) could fill-in the line center \citep{Hoard2003}.

The lower bound on the FWZI of J1923 is smaller than the FWZI of some high inclination SW Sextantis systems \citep{Dhillon2013}, which are thought to contain magnetically truncated disks \citep{Hoard2003}. This supports the idea that the accretion disk in J1923 may be truncated due to the presence of a magnetic field.

\subsubsection{Modelling the quiescent light curve} \label{Model_LC_Sec}
The light curve in quiescence was modelled using the Eclipsing Light Curve Code (ELC; \citealt{Orosz2000}). The basic model consisted of a WD with a temperature of 32000K (see Section \ref{Model_Spec_Sec} for more), an inner accretion disk radius in the range of 1-13$R_{WD}$, a temperature gradient in the disk ($T(r)=T_{i}\left(\frac{r_{i}}{r}\right)^{\xi}$) of 0.75 (which assumes a ``steady-state" disk), a hotspot with angular size in the range of 2-4$\degree$ centred on orbital phase 0.86, and an inclination between 72-90$\degree$. We computed best fit models based on the $\chi^{2}$ statistic using ELC for mass ratios (q=$\frac{M_{sec}}{M_{WD}}$) of 0.5, 0.4, 0.33, 0.285 and 0.25 and for inner disk temperatures ($T_{d}$) of 15000 K, 16000 K and 17000 K. Table \ref{ELCdata} shows the best fit inclination and $\chi^2$ for each model.

\begin{deluxetable}{ccccc}
\tablecolumns{5}
\tablewidth{0pc}
	\tablecaption{The best fit models computed using a $\chi^{2}$ grid search with ELC for various mass ratios (q) and inner disk temperatures ($T_{d}$).\label{ELCdata}}
	\tablehead{
		\colhead{q}	&\colhead{$T_{d}$(K)}	&\colhead{$i\degree$	}	&\colhead{$\chi^{2}$}	&\colhead{$\chi^{2}_{R}$}\\
		}
		
		\startdata
		0.5		& 15000			& 81.4-83.2 			& 2140 			&4.89\\
				& 16000			& 81.4-83.2 			& 2118 			&4.85\\
				& 17000			& 81.8-83.6 			& 2097 			&4.80\\
		0.4		& 15000			& 81.4-82.8 			& 2191 			&5.01\\
				& 16000			& 81.7-83.0 			& 2165 			&4.95\\
				& 17000			& 81.9-83.6 			& 2147 			&4.91\\
		0.33		& 15000			& 81.8-82.8 			& 2243 			&5.13\\
				& 16000			& 82.0-83.1 			& 2213 			&5.06\\
				& 17000			& 82.3-83.4 			& 2187 			&5.00\\
		0.285	& 15000			& 81.3-83.0 			& 2293 			&5.25\\
				& 16000			& 81.8-83.1 			& 2259 			&5.17\\
				& 17000			& 82.1-83.5 			& 2230 			&5.10\\
		0.25		& 15000			& 81.5-82.9 			& 2323 			&5.32\\
				& 16000			& 81.8-83.1 			& 2290 			&5.24\\
				& 17000			& 82.2-83.6 			& 2260 			&5.17\\
		\enddata
\end{deluxetable}

\begin{figure}
\epsscale{1}
\plotone{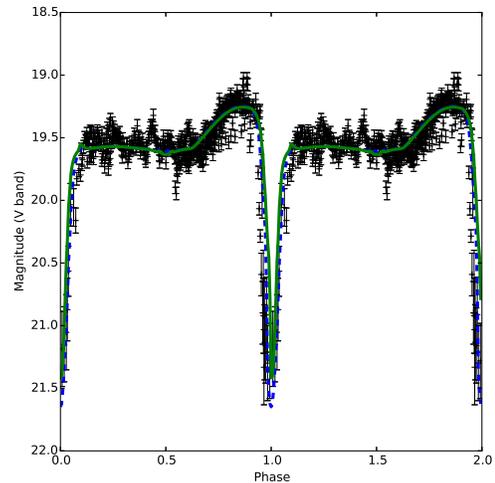}
\caption{The best (blue, dashed, q=0.5, T$_{d}$=17000 K, i=83.1\degree) and worst (green, solid, q=0.25, T$_{d}$=15000 K, i=82.25\degree) model light curves generated by ELC plotted against the low state data of J1923 (black). The ingress of the eclipse is deeper in the data than in any of our models, and is possibly related to how ELC models hotspots.}
\label{ELC_model}
\end{figure}

The resulting V-band fit to the data can be seen in Figure \ref{ELC_model}, which shows the best and worst fits of our modelling. The best fit inclination range over all models was very narrow, between 81.3$\degree$ and 83.6$\degree$. A mass ratio of q=0.5 provides the best overall fit.

The ingress of the eclipse is never well fit by any of our models, with the observed ingress always steeper than any of our models. This is a result of the simple model ELC assumes for the hotspot. ELC simply increases the temperature of the accretion disk cells which correspond to the position of the hotspot, instead of making the hotspot an independent structure. As such, ELC is not able to include eclipsing of the accretion disk by a physical hotspot for high inclination systems, which, in reality, would steepen the ingress.\\

\subsection{Modelling the spectrum} \label{Model_Spec_Sec}
We next modelled the GALEX FUV and NUV fluxes, along with the optical spectrum, to constrain the temperature of the WD and the contribution of the secondary star. WD model atmospheres were generated using Hubeny's TLUSTY, SYNSPEC and ROTIN programs (\citealt{Hubeny1988}; \citealt{Hubeny1995}) for WD temperatures between 24-34000 K and with a log $g$ of 8.

\cite{Schlafly2011} estimated the reddening in the direction of J1923 to be $E(B-V)=0.0471\pm0.0015$ mag. In the following, the spectrum has been de-reddened using this value and the Cardelli Extinction function \citep{Cardelli1989}, implemented in the \texttt{PYTHON} module \texttt{ASTROPYSICS}.

\begin{figure}
\epsscale{1.0}
\plotone{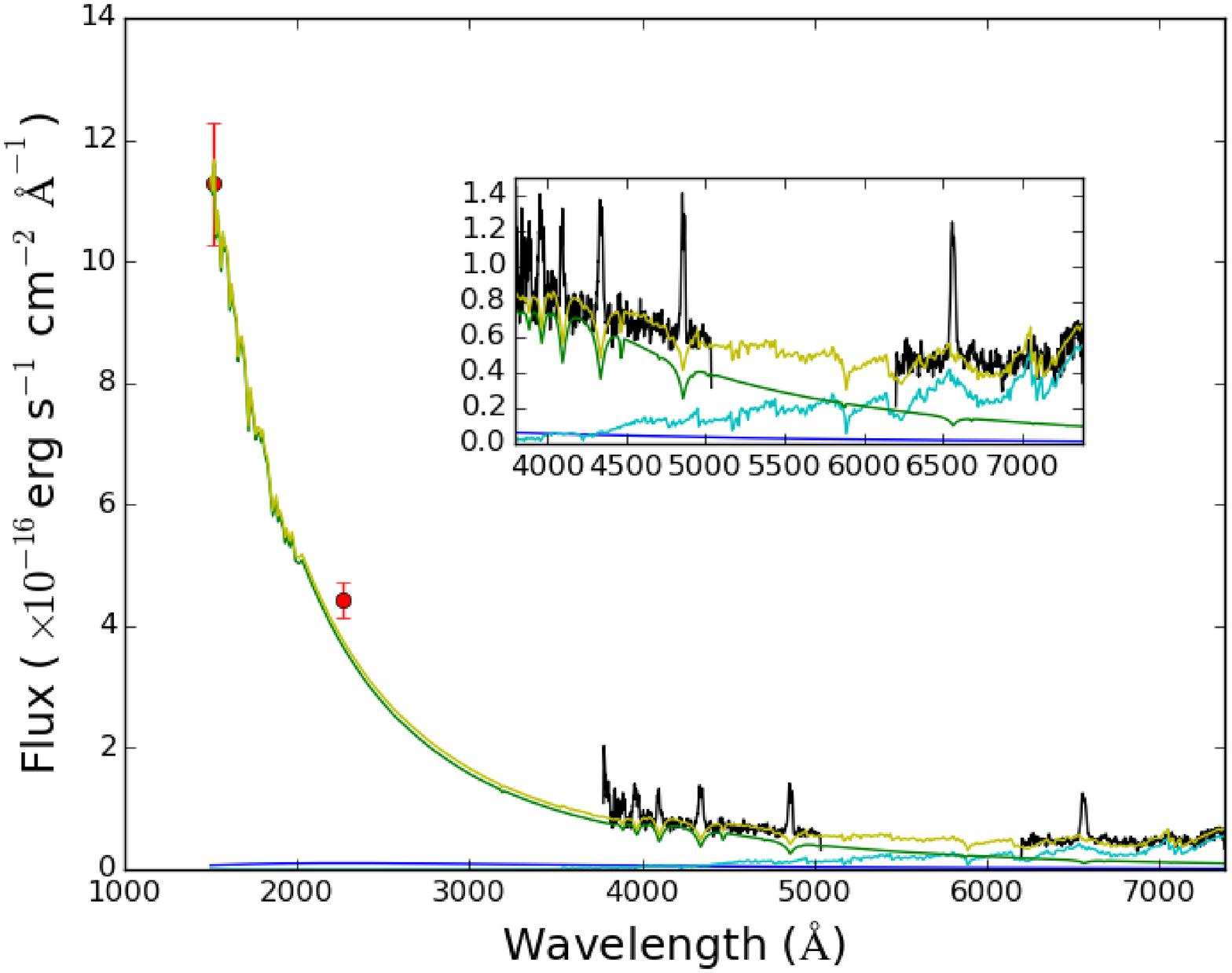}
\plotone{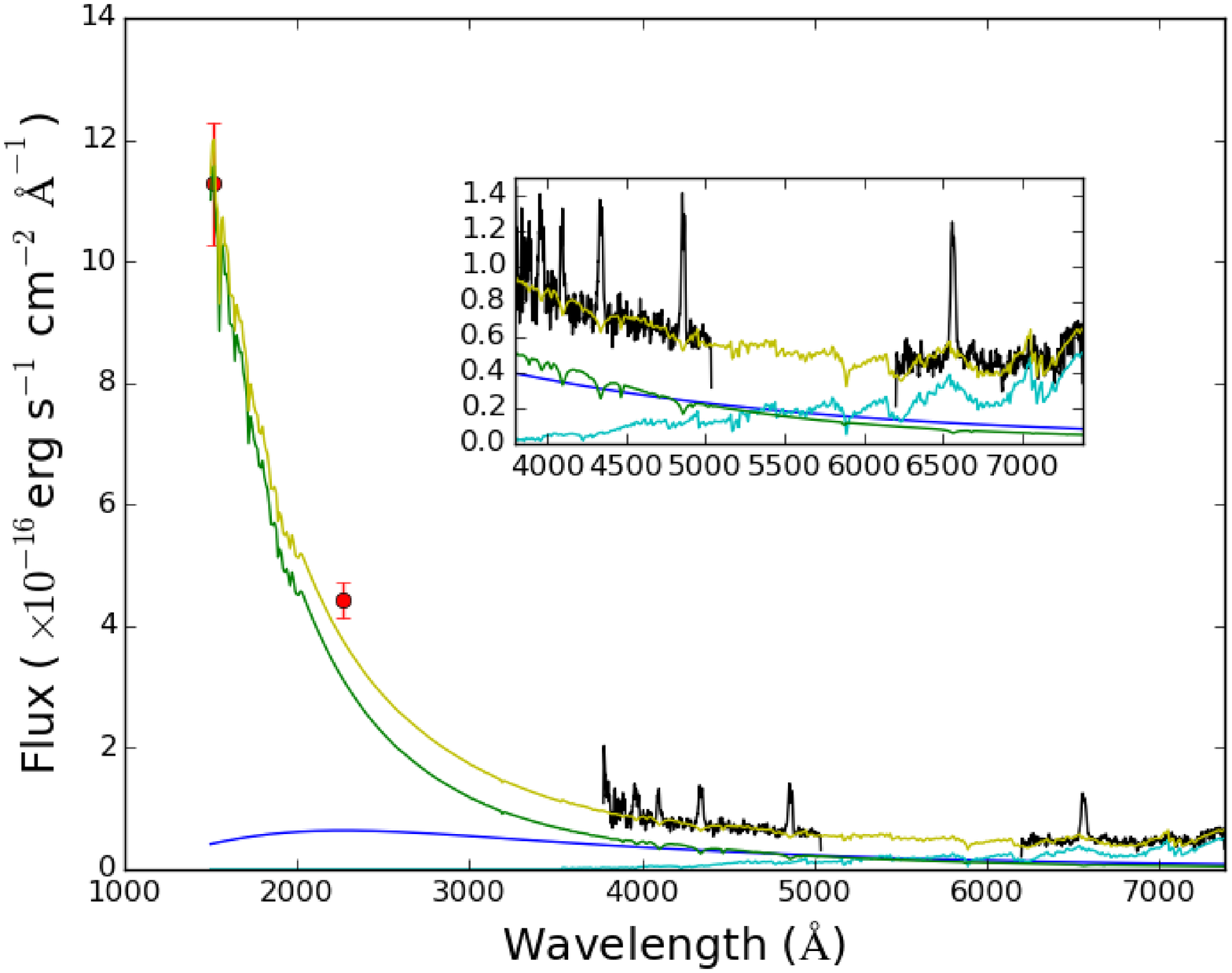}
\caption{\textit{Top:} The lowest WD temperature model spectrum fit to the GALEX fluxes and optical spectrum. The green curve is a WD model with T=24,000 K and log $g$=8. The green curve is an accretion disk with inner temperature of 15000 K and $\xi$=0.75. The cyan curve is the M5V spectrum from the MILES catalogue. The yellow is the sum of the three components. The inset is a zoom of the optical spectrum \textit{Bottom:} Requiring an equal contribution in optical flux between an accretion disk (with a maximum temperature of 15000K and a $\xi$=0.75, shown in blue) and a WD of temperature of T=34000 K gives a good fit to the spectrum, along with the same comparison star as in the top panel. A cooler WD allows for the WD to dominate in the optical, which is inconsistant with the results of the ELC modeling.}
\label{model_spectra}
\end{figure}

Table \ref{SpecFitdData} shows the results of our spectral fitting. The lowest temperature WD which provided an acceptable fit to the UV and the optical was a WD with a temperature of 24000 K and a minimum amount of disk contribution, along with a companion of spectral type M5V (the stellar spectrum was obtained from the MILES catalogue \citep{Peletier2006}). This fit can be seen in Figure \ref{model_spectra}.

\begin{deluxetable}{ccc}
\tablecolumns{3}
\tablewidth{0pc}
	\tablecaption{The reduced $\chi^{2}$  values for various fits to the optical spectrum and GALEX fluxes of J1923\label{SpecFitdData}}
	\tablehead{
		\colhead{WD Temp (K)}	&\colhead{$T_{d}$(K)}	&\colhead{$\chi^{2}_{R}$	}\\
		}
		
		\startdata
		24000		& 15000			& 1.15\\
					& 16000			& 1.15\\
					& 17000			& 1.16\\
		26000		& 15000			& 1.13\\
					& 16000			& 1.14\\
					& 17000			& 1.14\\
		28000		& 15000			& 1.12\\
					& 16000			& 1.12\\
					& 17000			& 1.13\\
		30000		& 15000			& 1.10\\
					& 16000			& 1.11\\
					& 17000			& 1.12\\
		32000		& 15000			& 1.09\\
					& 16000			& 1.10\\
					& 17000			& 1.11\\
		34000		& 15000			& 1.09\\
					& 16000			& 1.10\\
					& 17000			& 1.10\\
		\enddata
\end{deluxetable}

The best fit model from Table gives a WD temperature of 34000 K and an accretion disk with inner temperature of 15000 K. However, there is a degeneracy between all of our fits, which can be seen in the small change in the reduced $\chi^{2}$ between all of the models. The only limit our modelling reveals is that the lowest temperature WD which fits our data is 24000 K, as cooler WDs cannot be fit with an accretion disk contribution.

The contribution of the secondary is similar throughout our models, with a flux between $(2-2.5)\times10^{-16}$ erg s$^{-1}$ cm$^{-2}$ \AA$^{-1}$ at 5500\AA.

The V-band magnitude of our M5V template is 20.7$\pm$0.3. This gives a V-W1 of 4.1$\pm$0.3, assuming that the WISE magnitudes were taken when J1923 was in quiescence. This is much higher than the V-W1 of 6.2 for an M5V star, but is closer to the V-W1 of 5.2 for an M4V star \citep{Pecaut2013}. Hence, we conclude the companion spectral type is M4-5. This spectral type is also in keeping with the results from \cite{Knigge2011}, which shows that the expected spectral type for a CV with a period of 4 hours is a M4-M5. Furthermore, this spectral type is similar to the M5.5$\pm$0.5 companion of HS 0220+0603, another eclipsing SW Sextantis system with an orbital period between 3-4 hours \citep{Rodriguez-Gil2015}.

We use the V band magnitude of the companion in the model spectrum of 20.9$\pm$0.3, along with an absolute magnitude of $12.3$ for an M5 star, to compute a distance to J1923 is $d=520\pm80$ parsec. If the companion is a M4 star, then using an absolute magnitude of $11.2$, the distance to J1923 is $d=900\pm100$ parsec.

\subsection{Similarity to Lanning 386}
\begin{figure}[h!]
\epsscale{1}
\plotone{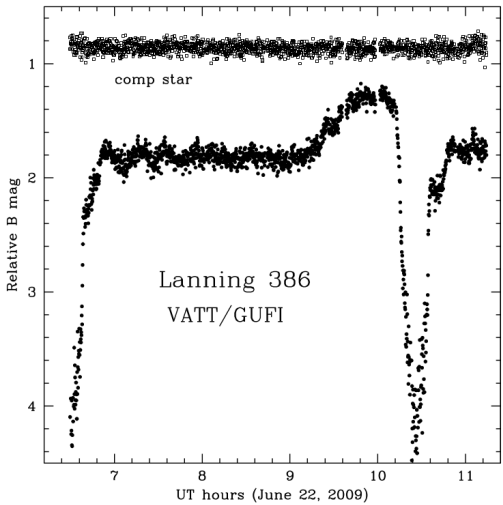}
\caption{Previously unpublished VATT/GUFI light curve for Lanning 386, showing pre-eclipse hump and QPOs}
	\label{new_lanning}
\end{figure}

Figure \ref{new_lanning} shows a previously unpublished light curve of Lanning 386 in quiescence, taken in 2009 using the Galway Ultra Fast Imager (GUFI) on the VATT. The calibrated V-band magnitude of Lanning 386 outside of eclipse is 17.5 in quiescence. The similarities between the light curve for Lanning 386 and those presented here for J1923 are beyond coincidence. Both objects show QPOs in the quiescent state, and have visible humps at phase 0.75. Both objects lose both of these features while in outburst, and brighten by similar magnitudes.

The spectra of both objects, taken when both were in quiescence, are also very similar. The Balmer decrements of Lanning 386 are $D_{34}=1.15$ and $D_{54}=0.85$ \citep{Brady2008}, which are within the measurement errors of the decrement for J1923. Both spectra also show the same M-type molecular band past 7200\AA. Lanning 386 displays 2 very different spectra, depending on its state. When in quiescence, its spectrum is identical to that observed here for J1923. However, in outburst, Lanning 386 displays prominent ionized helium emission lines, along with the Bowen Blend structure and C IV($\lambda=5806$\AA). The spectrum shown in Figure \ref{average_spectrum} was taken when J1923 was in the low state, which might explain the absence of the features seen in the excited state in Lanning 386.

The peak-to-peak separation of the emission lines in J1923 is small compared to the expected value for its high inclination, and the FWZI is smaller than the value typically found in SW Sextantis stars. We propose that this is possibly due to a truncation of the accretion disk due to the WDs magnetic field.

The single-peak structure seen in SW Sex stars is thought to arise due to a strong magnetic accretion curtain \citep{Hoard2003}. A possible explanation for the slightly double-peaked emission lines in J1923 could also be that there is weaker line emission from the accretion curtain in J1923, as opposed to the strong emission line core seen from the magnetic accretion curtain in SW Sextantis stars.

\section{Conclusions}
We have found the period of J1923 to be $P=0.16764612(5)$ day and determined a linear ephemeris for eclipses observed from April to June 2014. The best fit model to the eclipses requires a high inclination of 85.5$\pm$2\degree and a disk which does not extend to the surface of the WD. This inclination is greater than the inclination estimated using the peak-to-peak separation of the $H\alpha$ line. However, the small FWZI of the emission lines may suggest the accretion disk is truncated by the magnetic field of the WD, as proposed by \cite{Hoard2003}.

J1923 shares many similarities with Lanning 386. Both show QPOs in quiescence, have similar outburst amplitudes of $\Delta m\approx 2$, have similar orbital periods and have similar Balmer line spectra in quiescence. Even though it is still debatable whether both of these systems have anything to do with the SW Sextantis stars, it is nearly certain that they are extremely similar to each other. To confirm this, more simultaneous high-resolution spectroscopy and photometry of J1923 is required to see if the He II lines become prominent when the system is in outburst. It would also be of interest to see if the C IV 5806 \AA$\:$ line which is visible in Lanning 386 during outburst is also present in J1923, as the C IV line is a high excitation line rarely seen in CVs, yet has been observed in SW Sex, and is also present in Wolf-Rayet stars. It would also be interesting to observe J1923 in the X-ray region, as strong X-ray emission would suggest that J1923 may be a magnetic system. In that case, the accretion disk would be truncated before reaching the surface of the WD, which would help explain why the emission lines appear single-peaked, despite its inferred inclination. This scenario also requires a very weak emission component from a magnetic accretion curtain, as there is no evidence for a strong core in the emission lines of J1923.  If this X-ray emission were also seen in Lanning 386, it would help explain why the emission lines seem single peaked, despite the systems inferred inclination.

If J1923 does exhibit the He II line at 4686 \AA$\:$ when in outburst, then the question still remains as to what causes this emission in both J1923 and Lanning 386. The V Sagittae stars are CVs with long periods (5$<$P$<$24 hour) which display very prominent He II 4686 \AA$\:$ emission \citep{Steiner1998}. It is possible that the V Sagittae stars and SW Sextantis stars are related through stars similar to J1923 and Lanning 386, and that the He II 4686 \AA$\:$ emission is the key to linking these interesting classes together.

\section*{Acknowledgements}
\acknowledgments

MRK, PC and PM acknowledge financial support from the Naughton Foundation, Science Foundation Ireland and the UCC Strategic Research Fund. PS acknowledges support from NSF grants AST-1008734 and AST-1514737. We thank the Vatican Observatory and Richard Boyle for providing us observing time on the VATT. Some of this work is based on observations obtained with the Apache Point Observatory 3.5m telescope, which is owned and operated by the Astrophysical Research Consortium. We would also like to thank the anonymous referee for constructive and useful feedback.

{\it Facilities:} \facility{VATT}, \facility{ARC}, \facility{C2PU}.

\bibliographystyle{aasjournal}
\bibliography{j1923_kennedy.bib}

\begin{thebibliography}{}
\expandafter\ifx\csname natexlab\endcsname\relax\def\natexlab#1{#1}\fi

\bibitem[{Brady {et~al.}(2008)Brady, Thorstensen, Koppelman, Prieto, Garnavich,
  Hirschauer, \& Florack}]{Brady2008}
Brady, S., Thorstensen, J., Koppelman, M., {et~al.} 2008, Publications of the
  Astronomical Society of the Pacific, 120, 301

\bibitem[{Cardelli {et~al.}(1989)Cardelli, Clayton, \& Mathis}]{Cardelli1989}
Cardelli, J.~a., Clayton, G.~C., \& Mathis, J.~S. 1989, The Astrophysical
  Journal, 345, 245

\bibitem[{Dhillon {et~al.}(2013)Dhillon, Smith, \& Marsh}]{Dhillon2013}
Dhillon, V.~S., Smith, D.~a., \& Marsh, T.~R. 2013, Monthly Notices of the
  Royal Astronomical Society, 428, 3559

\bibitem[{Groot(2000)}]{Groot2000}
Groot, P.~J. 2000, New Astronomy Reviews, 44, 137

\bibitem[{Hoard {et~al.}(2003)Hoard, Szkody, Froning, Long, \&
  Knigge}]{Hoard2003}
Hoard, D.~W., Szkody, P., Froning, C.~S., Long, K.~S., \& Knigge, C. 2003, The
  Astronomical Journal, 126, 2473

\bibitem[{Horne \& Marsh(1986)}]{Horne1986}
Horne, K., \& Marsh, T. 1986, Monthly Notices of the Royal Astronomical
  Society, 218, 761

\bibitem[{Hubeny(1988)}]{Hubeny1988}
Hubeny, I. 1988, Computer Physics Communications, 52, 103

\bibitem[{Hubeny \& Lanz(1995)}]{Hubeny1995}
Hubeny, I., \& Lanz, T. 1995, The Astrophysical Journal, 439, 875

\bibitem[{Knigge {et~al.}(2011)Knigge, Baraffe, \& Patterson}]{Knigge2011}
Knigge, C., Baraffe, I., \& Patterson, J. 2011, The Astrophysical Journal
  Supplement Series, 194, 28

\bibitem[{Leach {et~al.}(1999)Leach, Hessman, King, Stehle, \&
  Mattei}]{Leach1999}
Leach, R., Hessman, F.~V., King, A.~R., Stehle, R., \& Mattei, J. 1999, Monthly
  Notices of the Royal Astronomical Society, 305, 225

\bibitem[{{Lomb}(1976)}]{Lomb76}
{Lomb}, N.~R. 1976, Astrophysics \& Space Science, 39, 447

\bibitem[{Orosz \& Hauschildt(2000)}]{Orosz2000}
Orosz, J.~a., \& Hauschildt, P.~H. 2000, Astronomy and Astrophysics, 364, 265

\bibitem[{Patterson {et~al.}(2002)Patterson, Fenton, Thorstensen, Harvey,
  Skillman, Fried, Monard, O'Donoghue, Beshore, Martin, Niarchos, Vanmunster,
  Foote, Bolt, Rea, Cook, Butterworth, \& Wood}]{Patterson2002}
Patterson, J., Fenton, W.~H., Thorstensen, J.~R., {et~al.} 2002, Publications
  of the Astronomical Society of the Pacific, 114, 1364

\bibitem[{Pecaut \& Mamajek(2013)}]{Pecaut2013}
Pecaut, M.~J., \& Mamajek, E.~E. 2013, The Astrophysical Journal Supplement
  Series, 208, 9

\bibitem[{Ritter \& Kolb(2003)}]{Ritter2003}
Ritter, H., \& Kolb, U. 2003, Astronomy and Astrophysics, 404, 301

\bibitem[{Rodr\'{\i}guez-Gil {et~al.}(2001)Rodr\'{\i}guez-Gil, Casares,
  Mart\'{\i}nez-Pais, Hakala, \& Steeghs}]{Rodriguez-Gil2001}
Rodr\'{\i}guez-Gil, P., Casares, J., Mart\'{\i}nez-Pais, I.~G., Hakala, P., \&
  Steeghs, D. 2001, The Astrophysical Journal, 548, L49

\bibitem[{Rodr\'{\i}guez-Gil {et~al.}(2007)Rodr\'{\i}guez-Gil, Gansicke, Hagen,
  Araujo-Betancor, Aungwerojwit, {Allende Prieto}, Boyd, Casares, Engels,
  Giannakis, Harlaftis, Kube, Lehto, Martinez-Pais, Schwarz, Skidmore, Staude,
  \& Torres}]{Rodriguez-Gil2007}
Rodr\'{\i}guez-Gil, P., Gansicke, B.~T., Hagen, H.-J., {et~al.} 2007, Monthly
  Notices of the Royal Astronomical Society, 377, 1747

\bibitem[{Rodr\'{\i}guez-Gil {et~al.}(2015)Rodr\'{\i}guez-Gil, Shahbaz, Marsh,
  G\"{a}nsicke, Steeghs, Long, Mart\'{\i}nez-Pais, {Armas Padilla}, Schwarz,
  Schreiber, Torres, Koester, Dhillon, Castellano, \&
  Rodr\'{\i}guez}]{Rodriguez-Gil2015}
Rodr\'{\i}guez-Gil, P., Shahbaz, T., Marsh, T.~R., {et~al.} 2015, Mon. Not. R.
  Astron. Soc, 452, 146

\bibitem[{Sanchez-Blazquez {et~al.}(2006)Sanchez-Blazquez, Peletier,
  Jimenez-Vicente, Cardiel, Cenarro, Falcon-Barroso, Gorgas, Selam, \&
  Vazdekis}]{Peletier2006}
Sanchez-Blazquez, P., Peletier, R.~F., Jimenez-Vicente, J., {et~al.} 2006,
  Monthly Notices of the Royal Astronomical Society, 371, 703

\bibitem[{{Scargle}(1982)}]{scargle82}
{Scargle}, J. 1982, The Astrophysical Journal, 263, 835

\bibitem[{Schlafly \& Finkbeiner(2011)}]{Schlafly2011}
Schlafly, E.~F., \& Finkbeiner, D.~P. 2011, The Astrophysical Journal, 737, 103

\bibitem[{Steiner \& Diaz(1998)}]{Steiner1998}
Steiner, J., \& Diaz, M. 1998, Publications of the Astronomical Society of the
  Pacific, 110, 276

\bibitem[{Stellingwerf(1978)}]{Stellingwerf1978}
Stellingwerf, R. 1978, The Astrophysical Journal, 224, 953

\bibitem[{Thorstensen {et~al.}(1991)Thorstensen, Ringwald, Wade, Schmidt, \&
  Norsworthy}]{Thorstensen1991}
Thorstensen, J.~R., Ringwald, F.~A., Wade, R.~A., Schmidt, G.~D., \&
  Norsworthy, J.~E. 1991, The Astronomical Journal, 102, 272

\bibitem[{Warner(1995)}]{Warner1995}
Warner, B. 1995, in Cataclysmic Variable Stars, ed. A.~King, D.~Lin, S.~Maran,
  J.~Pringle, \& M.~Ward (Cambridge: Cambridge University Press)

\end{thebibliography}

\end{document}